# A Large-Scale Study of a Sleep Tracking and Improving Device with Closed-loop and Personalized Real-time Acoustic Stimulation

**Authors:** Anh Nguyen[1]*, Galen Pogoncheff[3], Ban Xuan Dong[3], Nam Bui[2], Hoang Truong[2], Nhat Pham[4], Linh Nguyen[3], Hoang Huu Nguyen[3], Sy Duong-Quy[5,6,7], Sangtae Ha[2,3], Tam Vu[2,3,4]

**Affiliations:**

[1]Department of Computer Science, University of Montana, Missoula, Montana, USA

[2]Department of Computer Science, University of Colorado Boulder, Boulder, CO 80309, USA

[3]Earable Inc., Boulder, CO 80309, USA

[4]Department of Computer Science, University of Oxford, Oxford OX1 3QD, UK

[5]Penn State College of Medicine, Hershey Medical Center, PA, USA

[6]Lam Dong Medical College, Da Lat City, Lam Dong Province, Vietnam

[7]Pham Ngoc Thach University of Medicine, Ho Chi Minh City, Vietnam

*Corresponding author. Email: anh.nguyen@umontana.edu

**One Sentence Summary:** Our novel Earable system achieves high-quality biosignal acquisition, over 87% sleep monitoring accuracy, and over 24-min drop in falling asleep time.

**Abstract:**

Various intervention therapies ranging from pharmaceutical to hi-tech tailored solutions have been available to treat difficulty in falling asleep commonly caused by insomnia in modern life. However, current techniques largely remain ill-suited, ineffective, and unreliable due to their lack of precise real-time sleep tracking, in-time feedback on the therapies, an ability to keep people asleep during the night, and a large-scale effectiveness evaluation. Here, we introduce a novel sleep aid system, called Earable, that can continuously sense multiple head-based physiological signals and simultaneously enable closed-loop auditory stimulation to entrain brain activities in time for effective sleep promotion. We develop the system in a lightweight, comfortable, and user-friendly headband with a comprehensive set of algorithms and dedicated own-designed audio stimuli. We conducted multiple protocols from 883 sleep studies on 377 subjects (241 women, 119 men) wearing either a gold-standard device (PSG), Earable, or both concurrently. We demonstrate that our system achieves (1) a strong correlation ($0.89 \pm 0.03$) between the physiological signals acquired by Earable and those from the gold-standard PSG, (2) an $87.8 \pm 5.3\%$ agreement on sleep scoring using our automatic real-time sleep staging algorithm with the consensus scored by three sleep technicians, and (3) a successful non-pharmacological stimulation alternative to effectively shorten the duration of sleep falling by $24.1 \pm 0.1$ minutes.

These results show that the efficacy of Earable exceeds existing techniques in intentions to promote fast falling asleep, track sleep state accurately, and achieve high social acceptance for real-time closed-loop personalized neuromodulation-based home sleep care.

**Main Text:**

**INTRODUCTION**

Getting enough sleep is essential to health and wellbeing. However, it has been reported that hundreds of millions of people globally are getting insufficient high-quality shut-eye (*1*). According to the Centers for Disease Control and Prevention (CDC), an estimated 50 to 70 million people in the United States suffer from several sleep disorders such as insomnia, narcolepsy, and sleep apnea (*2*). Consequently, there have been many associated adverse outcomes, including heart disease and stroke (*3*), depression (*4*), and drug use (*5*). Hence, the Affordable Care Act set a goal to improve sleep health for Healthy People 2020, reflecting a growing trend among researchers to incorporate sleep in health promotion and disease prevention (*6*). Among several symptoms of sleep disorders, difficulty falling asleep is the most common trouble, highly ruins the whole sleep session at night, and significantly increases sleep debt (*2*). Further consequences include focus deficiencies, frequent headaches, daytime fatigue, mood disorders, low energy inducing severe degradation of next-day activities performance, and associated major mental diseases (*7*). Thus, treatment for such sleeplessness is critical to improving the quality of life physically and mentally.

Traditional treatments for trouble falling asleep consist of complementary medicine (*8, 9*), sleep supplements (*10*), and natural remedies (e.g., acupuncture, guided imagery, yoga, hypnosis, herbs (*11–13*)). Although drugs produce quick symptomatic relief, they are associated with short sustainability and multiple adverse side effects (*14*). On the other hand, the natural options are limited due to a need for therapists or high daytime consumption. Overcoming these limitations has introduced Internet-based CBT-i (ICBT-i) (*15*) as an advanced intervention given through a computer or a mobile device right at sleeping time. This approach mainly combines sleep restriction, stimulus control, cognitive restructuring, sleep hygiene education, and relaxation. Numerous studies demonstrated that ICBT-i had become an effective method due to its highly structured, content-specific, low-cost, and flexible therapeutic approach (*16*). Indeed, a systematic review of CBT-i (*17*) showed that the treatment significantly affects sleep onset latency (SOL), a crucial factor in evaluating the falling asleep difficulty level. However, despite its positive impact on counteracting sleep loss effects, ICBT-i has not been a fully automatic real-time solution. The most reason is that the therapy still goes on even after the user falls asleep, which negatively wakes them up or keeps their sleep shallow.

Recent advantages in bioelectric sensing technologies have fueled a growing brain-computer interaction (BCI) generation for medical diagnostics (*18*), personal health care (*19*), human-machine interfaces (*20*), etc. Specifically, the rise of sleep tracking technology has nowadays been a boon for sleep awareness. Beyond medical-grade devices, referred to as polysomnography (PSG) (*21*), there have been various hi-tech solutions developed to help assess sleep patterns, such as wristbands (*22–24*), headbands (*25–28*), ear-worn devices (*29, 30*), and smartphone apps (*31–33*), to name a few. Unlike PSG, these devices are usually compact, inexpensive, less burdensome, easily worn for multiple nights at home, and require minimal or no expert supervision. However, beyond comparing their performance of sleep monitoring to PSG, only a few help to address sleep loss issues, especially difficulty falling asleep, as desired with a limited number of studies. Consequently, it is not easy to independently verify their sleep solution's effectiveness.

This work introduces a novel sleep aid system, which we have named Earable, that can provide highly reliable, accurate, and effective physiological signal acquisition, automatic sleep stage analysis, and fast sleep promotion, respectively. In this work, we develop Earable in the form of a lightweight, comfortable, and user-friendly headband along with six dry active bioelectrodes, a 3-D accelerometer, and a pulse oximetry sensor. Through multiple unit tests, we first show that the headband for bioelectrical signal recording has both electrode-skin interface impedance and signal quality comparable to the clinical PSG. As a result, this validation illustrates the ability of Earable to monitor brainwaves and many essential vital signs, including heart rate, respiratory rate, and sleep postures, during sleep. Then, we demonstrate the power of Earable to track, promote, and improve sleep quality on a large scale. To this end, we use Earable and PSG simultaneously to conduct a comprehensive study in which we record the data from 377 subjects over 883 sleep sessions done in multiple study protocols. Leveraging such a big dataset, we show that Earable can perform a robust automatic sleep stage classification via a set of high-performance algorithms compared to a consensus of three (03) sleep technicians' manual scoring done with PSG data. We then prove the ability of Earable to deliver effective auditory stimulation on helping users fall asleep within a reasonable amount of sleep onset time. Conclusively, Earable proposes unique properties in high throughput, scalability, comfort, and cost-effectiveness, which highly support diverse applications in health care and human-computer interaction in the future.

## RESULTS

### Device design

Earable presents a complete wireless wearable system that incorporates several biopotential electrodes (or electrodes in short) and sensors in a single platform to aid sleep. Fig. 1A illustrates an overview of Earable in front and rear views. A key design objective of Earable is to collect many biosignals while eliminating all stiff sensing components in a sleep headband. Hence, we design the electrodes in different shapes and fabricate them with soft conductive materials. In addition, to further optimize the comfortability, we use an elastic band behind the head to make Earable a one-size-fits-all device by flexibly adjusting the headband tightness when worn on a user's head (Fig. 1B). Fig. 1C shows the integrated Earable platform, which incorporates eight dry active biopotential electrodes, an inertial measurement unit (IMU), a photoplethysmography (PPG) sensor, and two bone conduction speakers. Specifically, Earable uses stretchable biocompatible conductive fabric made of antioxidant silver-coated polyurethane for electrodes touching the skin on the forehead and above the ears. Additionally, to raise the reusability of these dry electrodes, we design a strip detachable from the mainboard to place them on for easy sanitization. The electrodes positioned on the hairy area, on the other hand, are made of flexible conductive silicone and consist of 6 prongs for maximizing hair penetration. As a result, we ensure better skin-electrode contact and a more stable signal during long wear. Therefore, all of them have the signal quality of commercial wet electrodes and the comfort and unobtrusiveness of dry electrodes to measure meaningful electrophysiological signals. Besides that, we use the IMU and PPG sensors to capture pulse waveforms from the head. We also embed two bone-conduction speakers inside the front of Earable to deliver auditory stimulation, which minimizes the sound that travels through air and affects their partner while keeping a perceptive sound loud enough for users. To further minimize users' sensitivity to hard components at sensitive pressured locations, increase its stretchability, and reduce its thickness, we inter-connect all electrodes to the device mainboard using flexible printed circuits as cables. Finally, we protect the mainboard, containing an integrated circuit, a Bluetooth low-energy (BLE) module, and a 500mAh, 3.7V lithium battery, from users' unconscious movements during sleep by placing it in a plastic box 3D-printed using very lightweight material. Using this hardware, we collect and stream the physiological signals to a host device (e.g., a smartphone), where we further deploy a pipeline of core algorithms to satisfy two critical goals: (1) tracking sleep stages precisely and (2) controlling the play of audio stimuli effectively in real-time to improve sleep quality. Fig. 1D provides the pipeline of our proposed algorithms, separated into two main components: (1) an Earable sleep staging model to process the raw data acquired through the BLE and infer the sleep stage, and (2) a personalized fast sleep stimulation to control the play of auditory stimuli automatically. We will describe all algorithm modules in detail in the following sections.

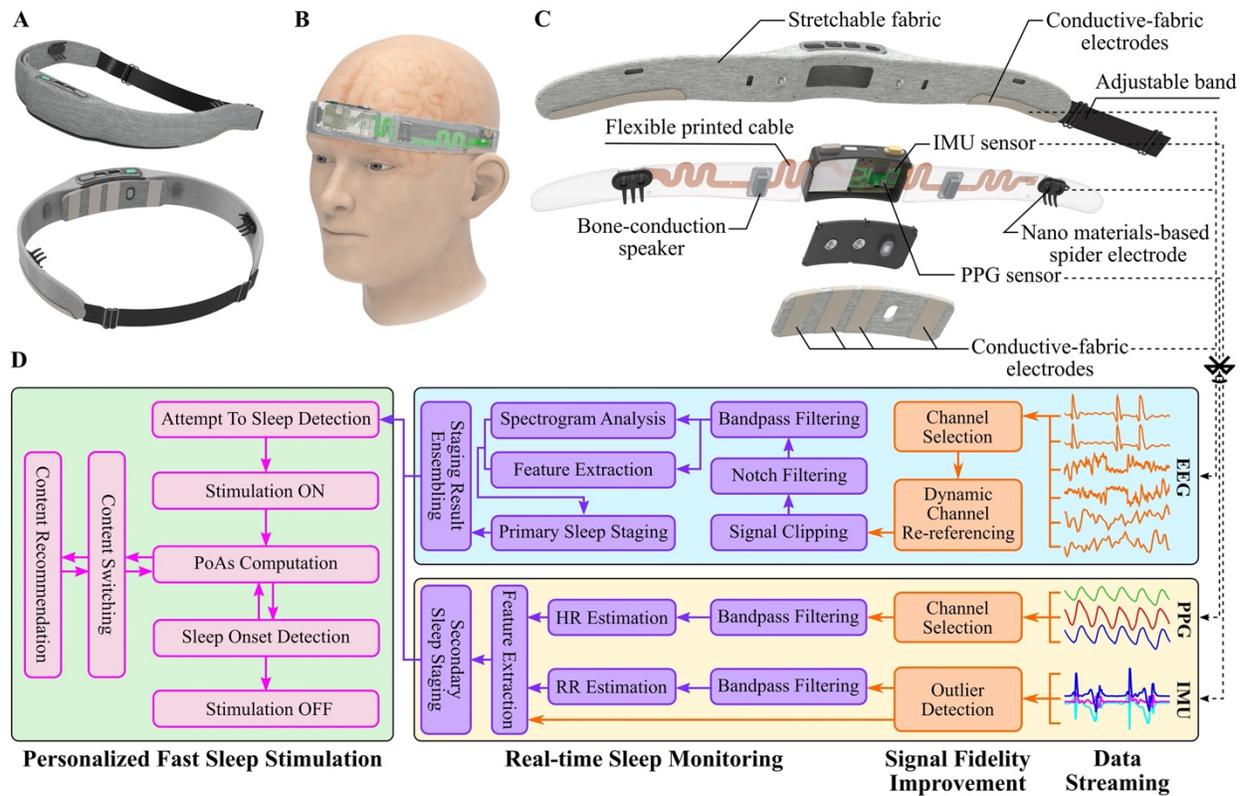

**Fig. 1. Earable overview.** (**A**) Device design. (**B**) Device placement on a user's head. (**C**) Exploded view. (**D**) End-to-end system pipeline. Head-based biosignals are collected with multimodal sensing hardware built in the headband and streamed via Bluetooth to a host device where several algorithms process the streaming data, extract sleep features, infer sleep stages, and activate and control auditory stimuli for real-time sleep assistance.

**Electrode montage and assessment of reliable signal acquisition**

We position the six bioelectrodes symmetrically around the head, including the two forehead (FH), two behind the ear (BE), and two over the ear (OTE) electrodes, close to the AF7, AF8, T9, T10, T7, and T8 positions, respectively, as defined by the international 10-20 system for electroencephalogram (EEG). The names of individual electrodes are specifically listed in Fig. 2A. In Fig. 2B, we provide the zoom-in images of both the fabric and conductive silicone materials mentioned in the previous section. Our electrode materials are highly conductive. Particularly, the surface resistance of conductive silicone is around $100 - 200\Omega.cm^{-1}$, whereas the surface resistance of our conductive fabric is around $1 - 2\Omega.cm^{-1}$ as measured by a bench-top multimeter.

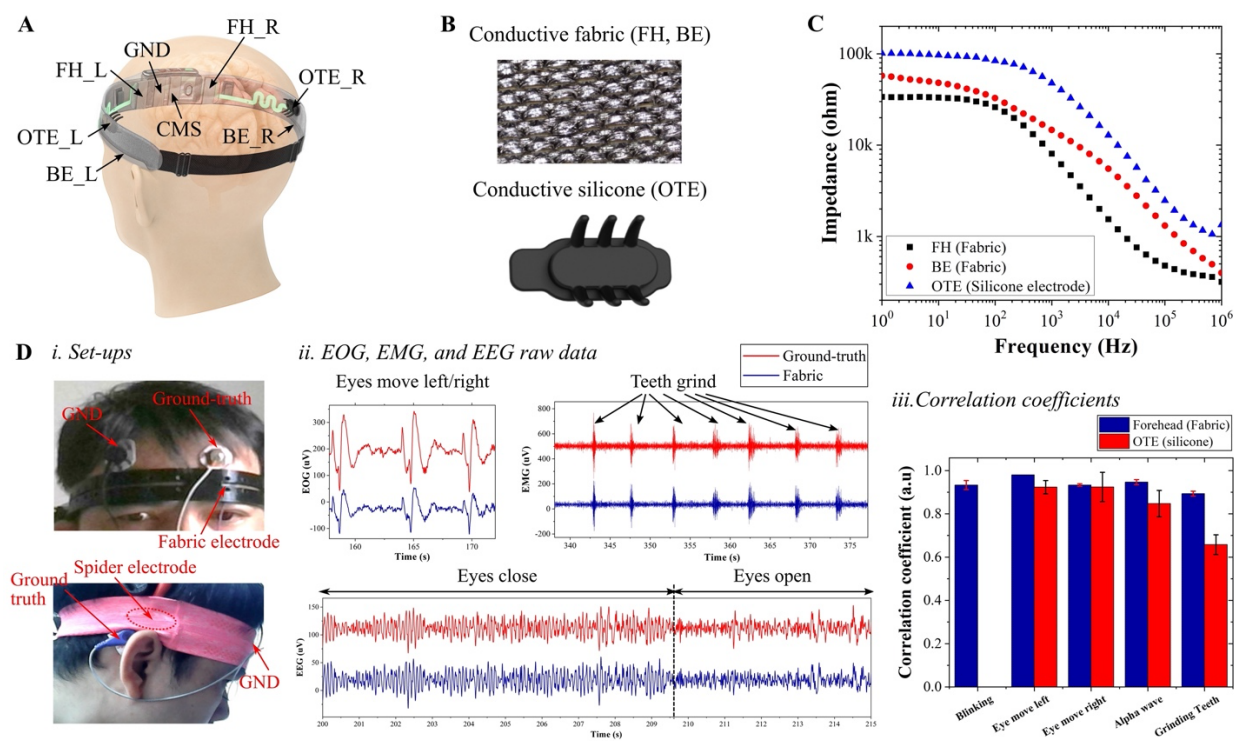

**Fig. 2. Electrode placement and electrophysiological recording using Earable.** (**A**) Name of biopotential electrodes located on the headband. (**B**) Micrograph of fabric and conductive silicone electrodes. The scale bar is 500µm. (**C**) Frequency-dependent impedance of FH, BE, and OTE electrodes. (**D**) i. Experimental setup for biopotential correlation tests for fabric FH (top) and conductive silicone OTE (bottom) electrodes. ii. Exemplary electrophysiological signals during several bio-calibration events collected from the FH location. Data is shifted vertically for clarity. iii. Correlation coefficients between electrophysiological signals collected using our dry electrodes and the clinical standard wet Ag/AgCl electrodes at the FH and OTE locations. Note that the eye blinking is not observable at the OTE location.

To evaluate electrode performance, we first measured electrode impedance using an electrochemical impedance spectroscopy (EIS). The measurement was done at three locations relevant to our electrodes. As demonstrated in Fig. 2C, from 1Hz to 1MHz, the impedances of three FH, BE and OTE electrodes are frequency-dependent and increase with reducing frequency. Notably, at all frequencies, the impedance of the FH electrode is lower than the BE one even though both electrodes are made from the same materials. This indicates that the electrode impedance is intimately tight to the location on the head from which the impedance is measured. The OTE electrode has the highest impedance, probably due to the small contact area available at the tip of its prongs. Since the impedance at each location was measured by connecting the electrode and reference to the two ends of the measuring device, it is reasonable to estimate each electrode's impedance as half of the measured values. Remarkably, at the low-frequency range of 1 – 100Hz, the impedance at all locations shows very weak frequency-dependent behavior, indicating a high contact quality between the electrodes and the skin so that their interface behaves like a pure resistor (*35*). Altogether, the high conductivity, low impedance, and intimate electrode/skin contact enable high-quality electrophysiological signal acquisition of our electrodes, as we will prove more in the following sections.

To demonstrate the functionality of our electrodes as an efficient electrophysiological signal monitor, we conducted experiments on tracking the eye blinking, eye moving left/right (EOG), teeth grinding (EMG), and alpha rhythms by closing eyes and relaxing (EEG) in comparison with signals collected by the ground-truth hydrogel Ag/AgCl electrodes. Fig. 2D shows the results of these experiments in detail. As illustrated in Fig. 2D(i), we performed two separate experiments at the FH and OTE locations by placing the Ag/AgCl electrodes near our dry electrodes to compare their signals. Exemplary EOG, EMG, and EEG signals collected at the FH location are shown in Fig. 2D(ii). For all events, the electrophysiological signals collected using our electrodes demonstrate striking similarity to those collected by the Ag/AgCl electrodes. They are able to obtain the low frequency (<4Hz) of the left/right eye movement as well as the high frequency (~100Hz) muscle activity during the teeth grinding experiment. Additionally, during the eyes closing and relaxing experiment, our dry electrodes are also able to capture the clear alpha activity occurring between 8 – 13Hz, which is not observable when opening the eyes. To further quantify the signal quality obtained by our electrodes, we computed the correlation coefficient between the signals collected from both electrode types as shown in Fig. 2D(iii). Overall, the two signals are strongly correlated as demonstrated by high correlation coefficients, in agreement with the raw data observed in Fig. 2D(ii). At the FH location, the correlation coefficients are $0.93 \pm 0.02$, $0.98 \pm 0.00$, $0.93 \pm 0.01$, $0.95 \pm 0.01$, and $0.89 \pm 0.01$ for eye blinking, eye moving left, eye moving right, alpha wave, and teeth grinding, respectively. At the OTE location, they are $0.92 \pm 0.03$, $0.92 \pm 0.07$, $0.85 \pm 0.06$, and $0.66 \pm 0.05$ for eye moving left, eye moving right, alpha wave, and teeth grinding (Note that at the OTE location the eye blinking is not observable). These values are slightly lower than at the FH location, which is probably due to the higher impedance of the conductive silicone electrode at the OTE location as shown earlier in Fig. 2C.

Having demonstrated the high fidelity of electrophysiological signals acquired using our dry electrodes, we proceeded to study the efficacy of our electrodes fully integrated with Earable in collecting full-night electrophysiological data compared with PSG. Fig. 3A shows a participant wearing our headband and hooked up to the PSG system in a sleep session. We then synchronized the data to perfectly align the two records. Beyond the headband data collected from 6 channels, the PSG data were collected from 8 standard locations in the 10-20 system, including F3/A2, F4/A1, C3/A2, C4/A1, O1/A2, O2/A1, LOC/A2, and ROC/A1. Fig. 3B depicts 8-second samples of electrophysiological signals recorded simultaneously from Earable's FH_L channel and PSG's F3/A2 channel. Comparable to prior results, the electrophysiological signals collected from our headband show great similarity to those collected from PSG in all four sleep stages. Earable is able to clearly capture sleep hallmarks traditionally observed in sleep studies such as alpha waves in the wake stage, spindles and K-complex in light sleep, and slow Delta wave in deep sleep (*36*).

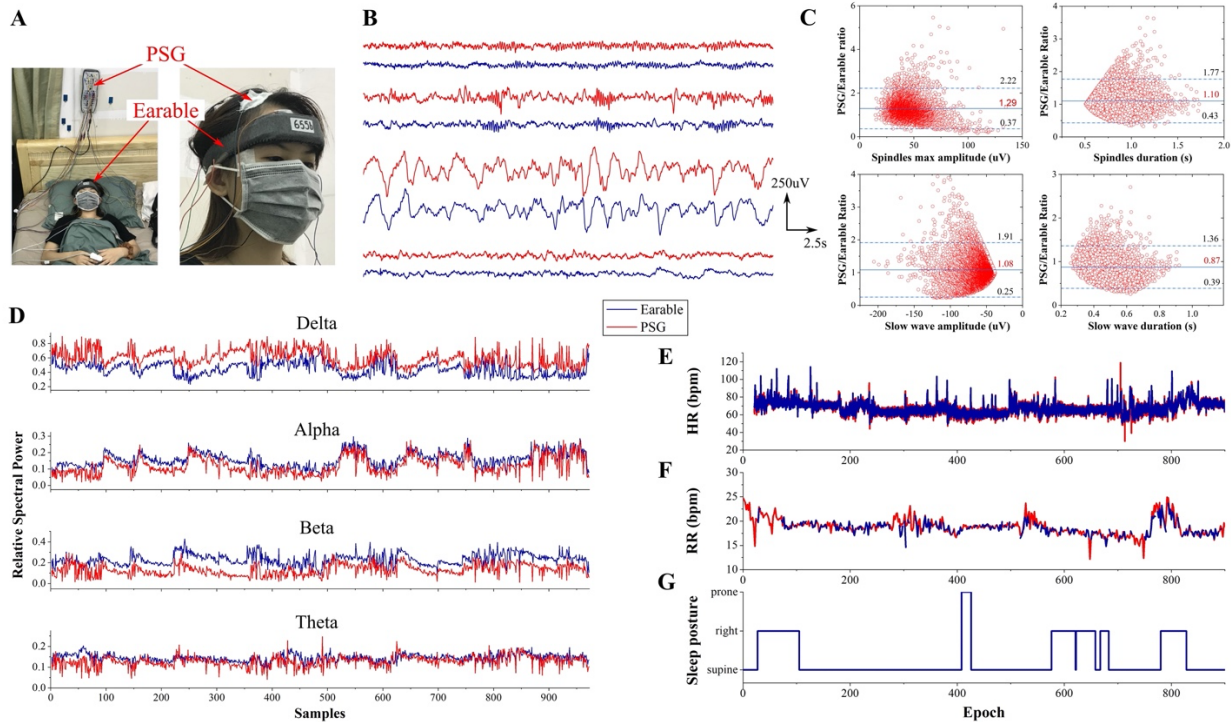

**Fig. 3. Signal quality using Earable in comparison with PSG in actual sleep studies.** (**A**) Clinical setup. A participant wears our headband and is hooked up to the PSG device at the same time for simultaneous validation of sleep microstructures and other vital signs. (**B**) Comparison of raw electrophysiological signals collected by two systems at four different sleep stages. (**C**) Bland-Altman analysis of slow waves and spindles. (**D**) Relative spectra power of Delta, Theta, Alpha, and Beta brainwaves collected during sleep. (**E, F**) Comparison of heart rate (E) and respiratory rate (F). (**G**) An example of a user's sleep posture throughout one full-night sleep.

To further quantify the data quality collected from Earable, we comprehensively performed the Bland-Altman analysis on the brainwave microstructures, as introduced by Leach et al. (*37*). Specifically, we compared important characteristics of slow waves and spindles collected from Earable and PSG, including the slow-wave duration, slow-wave negative amplitude, spindles frequency, and spindles max amplitude. We determined the duration of slow waves by computing the time from a negative peak to the next negative zero-crossing after that peak, whereas the maximum negative amplitude was the minimum amplitude of the signal during that time. For spindles analysis, we determined the spindle frequency as the number of positive peaks over the duration of a spindle event. We also calculated the spindle max amplitude as the maximum of absolute amplitude values during that time. These sleep microstructure characteristics were then compared by plotting the mean value against the ratio between the values done by two measurements, the results of which are shown in Fig. 3C. For each measured parameter, we show on the graph the mean difference between the two paired measurements (solid blue lines) and the 95% limits of agreement (dashed blue lines). Overall, we found that the sleep microstructures measured by both systems were similar. The relative limits of agreement of spindle max amplitude, spindle duration, slow-wave amplitude, and slow-wave duration are from 0.37 to 2.22, 0.43 to 1.77, 0.25 to 1.91, and 0.39 to 1.39, respectively. Note that our agreement values are not so high as those reported by Leach et al. (*37*), probably because the placement of our electrodes on the device is not the same as in the PSG montage, resulting in slight amplitude differences collected. Notably, we found that, on average, the amplitude of spindles and slow waves measured by PSG is about 29% and 8% higher, respectively, than those measured by our device. However, as shown in the next sections, these differences do not affect the high fidelity of the signals and sleep scoring capability of our system.

To compare with the results reported in previous studies (*38, 39*), we also compute the relative spectral power (RSP) in relevant frequency bands for sleep monitoring, including Delta (0.5 – 4Hz), Theta (4 – 8Hz), Alpha (8 – 13Hz), and Beta (15 – 30Hz), to assess the headband capacity in monitoring specific brainwaves during sleep. Fig. 3D demonstrates the RSP from these brainwave samples collected using Earable and PSG. Although the values from both systems are not strictly equal, they follow the same trend as the participant goes through different sleep stages. As a measure of agreement of the RSP between the headband and PSG, we further computed the mean absolute error (MAE) and the Pearson correlation at each frequency band. For Delta, Theta, Alpha, and Beta bands, the MAE values are 0.1636, 0.0213, 0.0402, and 0.1013, and the Pearson correlation values are 0.7030, 0.5026, 0.8794, and 0.4877, respectively. These values indicate a good agreement between the electrophysiological signals collected by Earable and PSG throughout the whole night sleep session.

Not only the brainwaves, but we also collect the heart rate (HR), respiratory rate (RR), and sleep positions to demonstrate the full capability of our headband in describing the whole spectrum of sleep data. We calculated the HR from the PPG sensor located next to the CMS electrode, whereas computing the RR and sleep position from our integrated IMU sensor. Details of the HR and RR calculations are presented in Supplemental Materials. Figs. 3E and 3F depict the HR and RR extracted from our headband compared to the ones extracted from gold standard EKG data and breathing belt connected to the PSG, respectively. As seen in the two sub-figures, the HR and RR extracted from our device follow almost exactly the ground-truth data. In particular, the MAE is $2.51 \pm 1.29$ bpm for HR (over nine studies) and $1.01 \pm 0.72$ breaths/min for RR (over 52 studies). These results indicate the high accuracy of Earable in capturing not only electrophysiological data but also other vital information during sleep. Finally, as exemplified in Fig. 3G, our device is also able to infer the participant's sleep posture using the IMU sensor.

**Real-time sleep scoring performance in overnight sleep studies**

Figure 4A depicts the 4-stage hypnograms from a trained sleep technician and the Earable automatic sleep scoring algorithm. In a typical night of sleep, people spend approximately 55% of total sleep time in light sleep (NREM stages 1 and 2), 20% in deep sleep (NREM 3), and 20-25% in REM (R) (*40*). During a sleep session, the wake stage typically presents itself prior to sleep onset, after the morning wake-up, and sporadically throughout the night. Given this imbalanced distribution of classes, we evaluate the metrics of precision, recall, and F1 scores for each of these 4-classes. We then use multi-class accuracy to summarize the agreement between the Earable sleep scoring algorithm output and the consensus scoring. Specifically, we randomly withheld 49 sleep studies from the development and training process of the primary machine learning (PML) model (described later in Supplemental Materials) for evaluation. As averaged over each test-set sleep study, we achieved a scoring accuracy of 87.8 ± 5.3%. F1-scores of 0.90 (precision: 0.91, recall: 0.88), 0.87 (precision: 0.87, recall 0.87), 0.91 (precision: 0.90, recall 0.91), and 0.82 (precision: 0.81, recall 0.84) were achieved for the wake, light sleep, deep sleep, and REM stages, respectively. In Fig. 4B, we provide a confusion matrix summarizing model performance behavior on this test-set. In addition, it is well-known that a Cohen's kappa (*41*) has been widely calculated as a reliable metric to avoid the inevitable variability and disagreement between different ways of sleep scoring. The Cohen's kappa of 0.8 or higher is considered as a strong agreement once we take into account the variabilities (*42*). For our study test set, we achieved the Cohen's kappa coefficient of 0.83 between the Earable sleep scoring algorithm output and the consensus scoring. For comparison, the Cohen's kappa coefficient between the scoring manually done by each sleep technician and the consensus scoring was 0.92 on average.

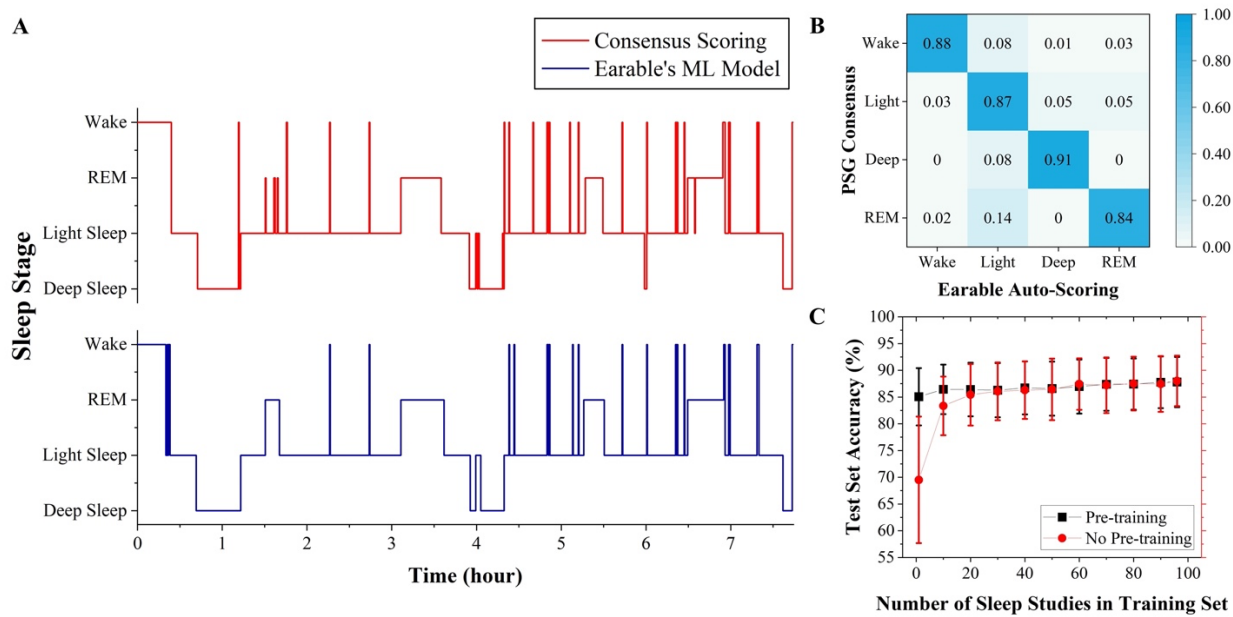

**Fig. 4. Accuracy performance of the Earable sleep stage classification model in overnight sleep studies.** (**A**) Comparison of hypnograms inferred from the Earable model and scored by a sleep technician in one study. (**B**) Confusion matrix. (**C**) Impact of training dataset size on the PML model performance.

On the other hand, the sleep data from a wide variety of users are usually required to develop a generalizable sleep scoring model that will perform well regardless of inter-user signal variability. Hence, Fig. 4C illustrates model performance as a function of training set size when using the PML model with pre-trained weights and the same model with randomly initialized weights. When training data is limited to a small number of sleep studies (i.e., ≤ 20), the influence of pre-training becomes significant in mitigating the PML model overfitting. As more studies become available for training, we can observe similar test-set performance between the pre-trained and randomly initialized PML models. Moreover, test-set accuracy increases as the size of the training dataset continues to grow but reaches a point where the average scoring accuracy among all users of the test set stops increasing with statistical significance. The arrival at such a plateau is indicative that a sufficient training dataset size is available, given the PML model architecture and the test dataset.

**Efficacy of channel selection and Significance of dynamic re-referencing scheme**

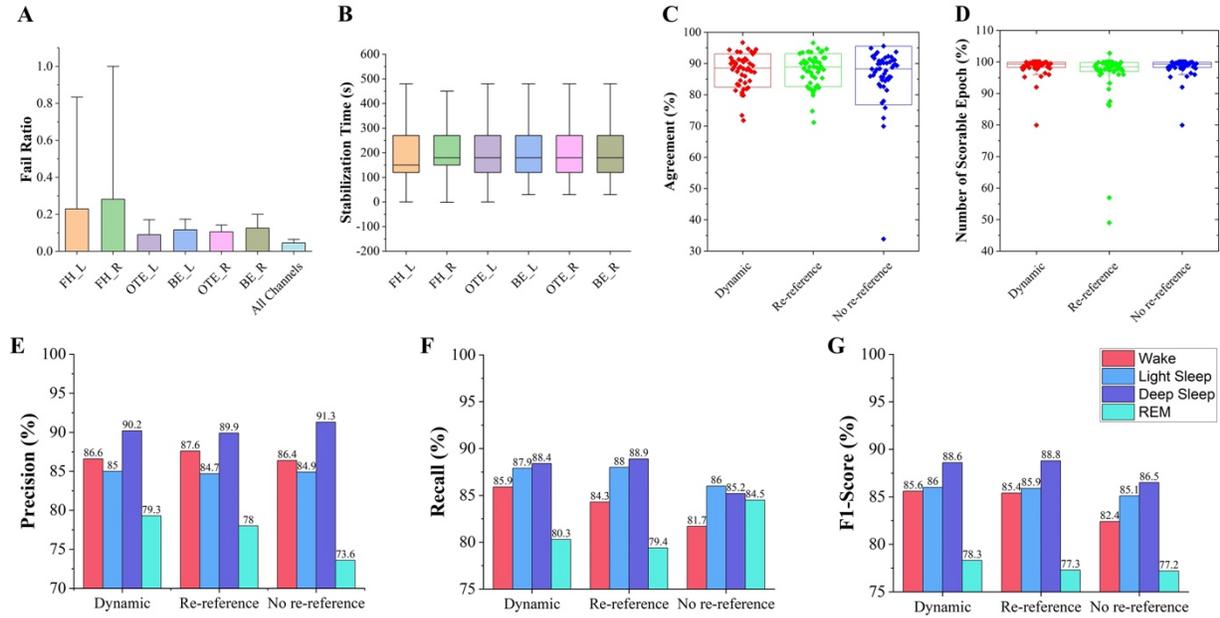

**Fig. 5. Accuracy performance of the Earable sleep stage classification model validated in modules.** (**A-B**) Signal quality validation for individual channels in terms of fail ratio (A) and stabilization time (B). (**C-G**) Accuracy performance in various channel re-referencing schemes.

Many automated sleep scoring algorithms rely on the availability of multiple bioelectrode source locations to enable accurate scoring. In at-home use cases, electrode contact quality and impedance are highly dependent on users' diligence and may be disturbed by their movements. Consequently, it becomes unlikely that multiple, pre-specified channels will be consistently available for sleep staging. As a result, we first develop a Channel Selection module (described later in Supplemental Materials) to identify which data channels contain a high enough fidelity signal for sleep stage classification. Fig. 5A depicts the proportion of times in which epoch data from each source channel are rejected by applying this module. Through the evaluation of the 49 subjects of the held-out test set, we observe that the OTE_L channel has the greatest availability with a minimal fail rate of 4.59%, followed by OTE_R (6.55%), BE_L (7.30%), BE_R (8.22%), FH_L (19.12%), and FH_R (24.96%) electrode derivations. Specifically, the electrophysiological data was unscorable only 1.58% of the time when considering all electrode derivations together. Upon review of videos that we simultaneously recorded during data acquisition, the most common reason for channel unavailability was the movement of the headband itself on the users' heads. In some cases, this movement resulted in improper electrode-skin contact at specific channels, causing these channels to become unavailable for long periods after the movement. In other cases, the movement only caused a temporary loss of contact while the electrode was momentarily shifted on the user's head, in which case the signal would quickly restabilize, probably due to the highly conformal electrode materials (conductive fabric and conductive silicone that could promptly establish stable contact with the user's skin.) (Fig. 5B).

To further deal with such a sudden change of channel data, we develop the Earable sleep scoring algorithm to be channel-location agnostic through a Dynamic Re-referencing module (described later in Supplemental Materials). It means that the PML model can accurately infer sleep stages regardless of input channel location, making it more robust to these potential issues. So, if more than one channel has good signal quality at a given time, the sleep stage probability estimates are aggregated from each channel to achieve more confident sleep stage predictions. To evaluate the efficacy of this module, we evaluate the Earable sleep scoring algorithm in the context of three different channel referencing schemes, including

- Scheme 1: Dynamic re-referencing
- Scheme 2: Re-referencing, where each FH, OTE, and BE channel is re-referenced to the BE channel on the opposite side of the headband.
- Scheme 3: No re-referencing, in which all source electrodes are referenced to the CMS electrode.

Boxplots in Fig. 5C summarize the distribution of 4-stage scoring accuracies (i.e., agreement) over the 49 subject test set when each scheme is applied. We achieve the greatest accuracy when the electrodes were always re-referenced to the appropriate BE channel. However, a one-way ANOVA (Analysis of Variance) test reveals that the mean accuracy scores across schemes are not significantly different at the 0.05 significance level ($p^* = 0.388$). Although, the re-referencing scheme is least robust to signal noise, as measured by the fraction of scorable epochs (i.e., epochs with at least one scorable signal channel) (Fig. 5D). On the other hand, we observe a significant difference in the percent of scorable epochs between these three schemes, with 98.42% of epochs evaluated as scorable using dynamic re-referencing and no re-referencing and 96.47% of epochs evaluated as scorable when applying re-referencing ($p^* = 0.039$). Although dynamic re-referencing and no re-referencing schemes enabled a greater percentage of epochs to be scored in this testing data, posthoc tests showed that these differences were not statistically significant ($p^* = 0.068$). Figs. 5E-G further depict per-stage precision, recall, and F1 scores, respectively. As the three approaches have the same statistical mean accuracy, the dynamic referencing scheme overperforms the other schemes in terms of flexibility.

**Scoring in the absence of high-fidelity electrophysiological signals**

While scoring the sleep data, there are epochs during which none of the electrophysiological channels have sufficient quality for sleep scoring (i.e., the epoch is "unscorable"). However, as our system still needs the real-time scoring result to control the closed-loop stimulation, other sensing modalities are utilized to substitute the unscorable channels. We targetted biosignal information extracted from the IMU and PPG sensors for scoring inference. Therefore, we build a secondary machine learning (SML) algorithm that uses the information derived from heart rate, respiratory rate, and movement estimates from the PPG and IMU sensors (refer to Supplementary Material for details regarding individual estimation). We conducted the evaluation using a nine-session SML model test-set and report the result of this approach as Case 1 in Fig. 6. We observe that the sleep scoring accuracy of the SML model is at 45.57%, as shown in Fig. 6E, with per-class F1 scores of 41.66%, 37.71%, 59.84%, and 47.20% for stages Wake, Light Sleep, Deep Sleep, and REM, respectively. Fig. 6A demonstrates that the SML model enables more epochs to be scored than when applying the PML model alone.

Beyond that, we also develop a model of rule-based scoring decisions to deal with when the epochs originally were not scorable using both the PML and SML models in real-time. Using a Hidden Markov Model (HMM) based smoothing algorithm, this offline model further helps reduce the number of unscorable epochs when users review a historical hypnogram shown after sleep. In a two-pass process, for each unscored epoch, the stage of the epoch is first estimated using a small set of rules following scoring manual procedures and statistical heuristics guided by the American Academy of Sleep Medicine (AASM) (*40*). Next, the Viterbi algorithm (*43*) is performed over the sequence of all inferred sleep stages of the hypnogram to compute the most likely sequence of stages in the hypnogram. If a more likely sequence of sleep stages is found in this algorithm, the hypnogram of the user is adjusted accordingly. This not only enables scoring of originally unscored sleep stages (permitting 100% of epochs to be scored in this test dataset (Fig. 6A) but also helps remove unlikely sleep stage transitions that were inferred by the PML and SML models when processing abnormal signals. HMM transition and emission probabilities were statistically computed from ground truth training data hypnograms (*44–50*).

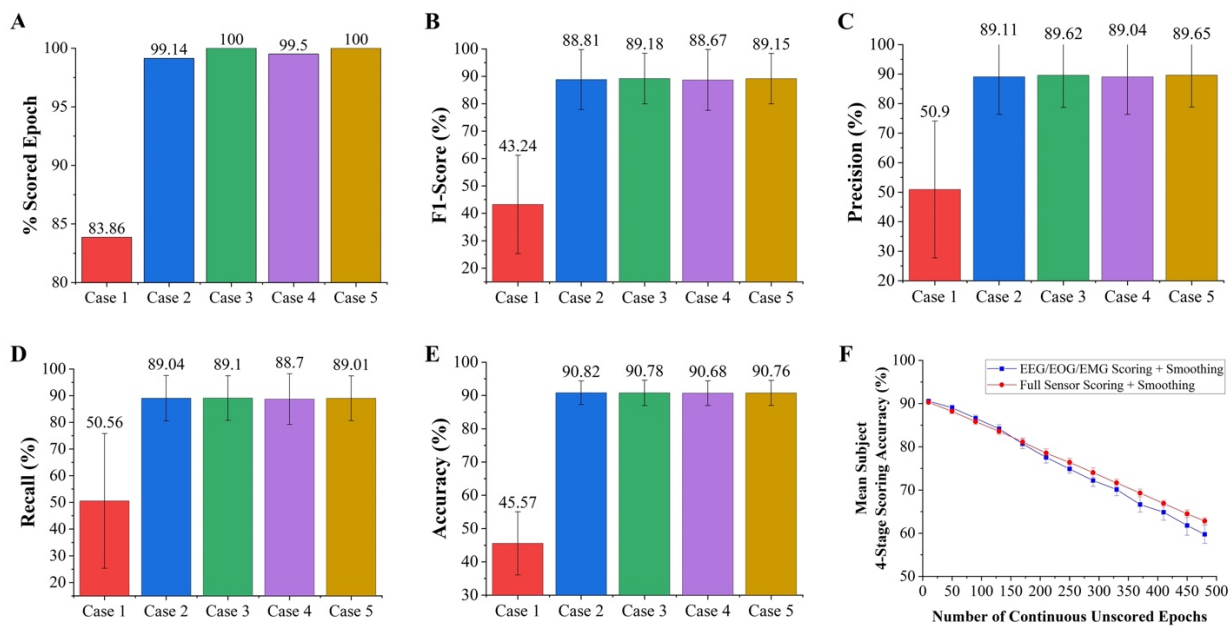

**Fig. 6. Further improvement of sleep scoring via smoothing and full sensor.** (**A-E**) Case 1: Scoring with IMU and PPG only. Case 2: Scoring using only ExG (including EEG, EOG, and EMG) data with no smoothing. Case 3: Scoring using only ExG data with HMM smoothing. Case 4: Scoring using all sensors with no smoothing. Case 5: Scoring using all sensors with HMM smoothing. (**F**) Scoring in the Presence of Long Poor Quality EEG Signals.

To understand the performance when scoring sleep using PML, SML, and offline smoothing models when applied to a 9-session test set (the intersection of the PML and SML model test sets), Figs. 6B-E show the accuracy, precision, and recall values, respectively, under the five following cases.

- Case 1: Scoring with the SML model only (ignoring sleep windows with excessively noisy PPG or IMU data)
- Case 2: Scoring with the PML model only (ignoring epochs that were left unscored by the PML model)
- Case 3: Scoring with the PML model and HMM smoothing.
- Case 4: Scoring with the PML and SML models.
- Case 5: Scoring with the PML model, the SML model, and HMM smoothing.

In a one-way ANOVA test, we observe no statistically significant difference in the accuracy, macro-precision, macro-recall, or macro-F1 between Case 2-5 ($p=0.9999$, $p=0.9979$. $p=0.9989$, and $p=0.9963$, respectively). In Cases 3 and 5, however, our proposed system scores every epoch in the test dataset, demonstrating the retention of accuracy after scoring originally ignored epochs.

Given the statistically similar accuracies between Case 3 and Case 5, it may be presumed that the SML model is unnecessary in this algorithm pipeline. However, removing the SML model will lead to a limitation in the Offline Smoothing algorithm, which exists in the fact that inference becomes less accurate as the duration of an unscorable segment increases. This limitation results from less local context for rule-based decisions and a decreasingly confident path in the Viterbi algorithm. This behavior is demonstrated by performing an experiment in which segments of increasing duration were synthetically removed from estimated hypnograms of the test set. Specifically, we first remove contiguous segments ranging in duration from 5 minutes (10 epochs) to 4 hours (480 epochs) from randomly selected times in each of the estimated hypnograms. Next, in one scenario, we apply the SML model to score these missing epochs using only data from the IMU and PPG, while in an alternate scenario, we leave these unscored epochs as is. In both scenarios, we finally apply the offline smoothing algorithm to score any remaining unscored epochs. The accuracy results of this process were averaged over 50 trials to reduce the variability in performance that may have resulted from where in the sleep session the segment was removed from. As apparent in Fig. 6F, the SML model becomes advantageous to offline scoring performance if the duration of electrophysiological unavailability is greater than 75 minutes (150 epochs).

**Sleep faster program performance via real-time closed-loop audio stimulation**

The sleep onset latency (SOL) is defined as the time when a user undergoes the transition from wake to other sleep stages such as light sleep, deep sleep or REM. In a normal, healthy adult, the SOL is usually in the 10 – 20 min range (*40*). In this section, we demonstrate that our Earable headband is capable of reducing SOL via real-time closed-loop audio stimulation, thereby helping users fall asleep faster. Specifically, for our SOL reduction protocol, we choose users that experiences difficulty in falling asleep fast (SOL > 20 min). First, to demonstrate that Earable is capable of accurately detecting SOL, we show in Fig. 7A the deviation of SOL measured by the Earable system from the ones measured by PSG. The result shows that in the 49 test-set studies, 96% of them have SOL estimated by Earable within ± 5 epochs (2.5 min) of the one extracted from PSG, indicating the high accuracy of our Earable headband in detecting SOL. Next, we use a well-structured audio stimulation input built using the structure shown in Fig. 7B to produce the best audio quality for fast sleep stimulation. Distinctively, we custom-build the audio files consisting of a guided breathing voice (GBV), a relaxation therapy voice (RTV), and background music (BM). The audio volume for each voice is faded out at each appropriate timeline as indicated in Fig. 7B to produce the best sleep promotion effect. The GBV is to help users breathe at appropriate paces, the RTV is to help users relax, and the BM is to help users feel relaxed and hence fall asleep easier. As illustrated in Fig. 7B, at t0, we turn on the GBV to help the user breathe at a regular pace while slowly turning on the BM. From t1 to t2, we apply the RTV music to help the user relax and start falling asleep. From t2 onward, only the BM remains. Depending on the sleeping protocols described below, the BM is automatically turned off after 50 minutes or will slowly fade out upon detecting the sleep onset latency (SOL).

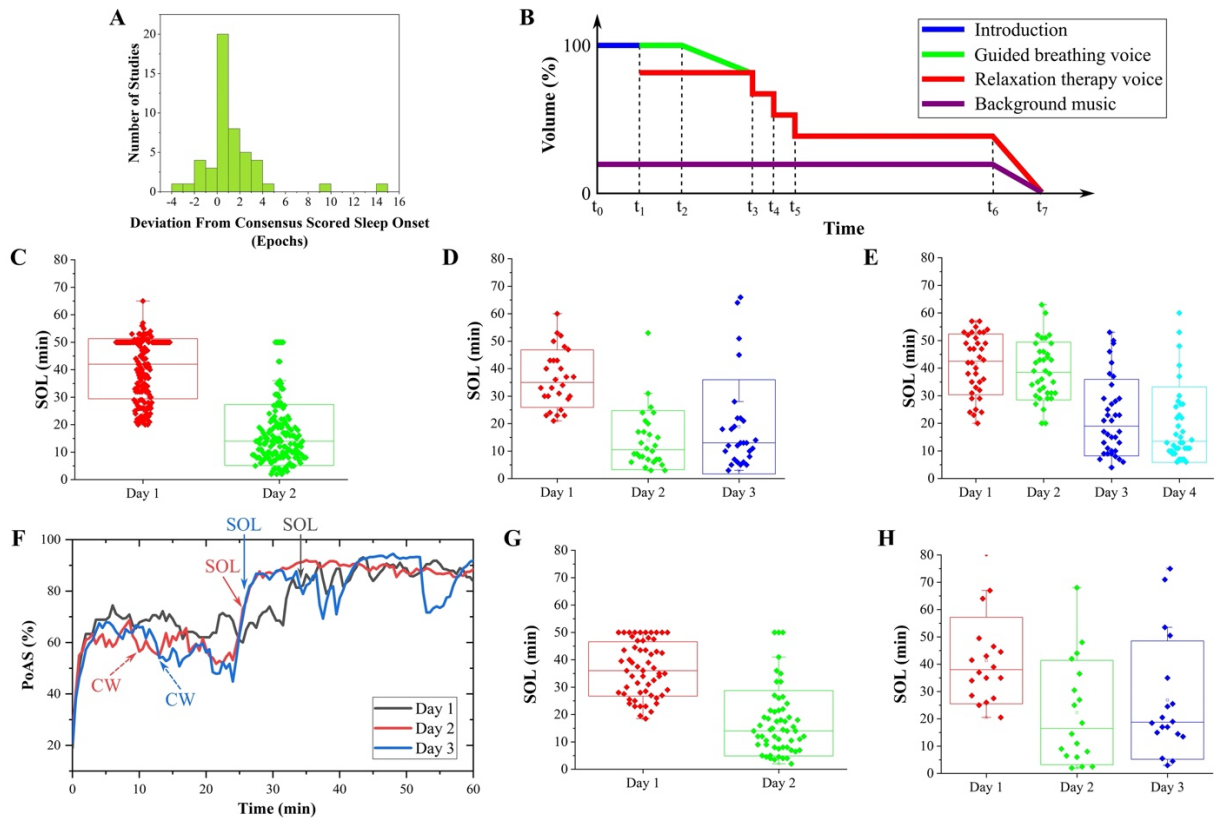

**Fig. 7. Performance of our proposed sleep faster program.** (**A**) Statistical distribution of deviation of sleep onset latency (SOL) estimation between Earable and PSG. (**B**) The structure of stimulation audio/voice used in our stimulation protocols. (**C**) SOL results of the 2-day nap protocol using PSG. Stimulation is used on Day 2 and not used on Day 1. (**D**) SOL results of a 3-day nap protocol using PSG. Stimulation is used on Day 2, 3 and not used on Day 1. (**E**) SOL results of the 4-day nap protocol using PSG. Stimulation is used on Day 3, 4 and not used on Day 1, 2. (**F**) Example of Probability of Being Asleep (PoAs) of a user at the beginning of 3

sleep sessions Day 1, 2, and 3. Closed-loop stimulation is used on Day 2 and 3. The solid arrows indicate the SOL measured by sleep staging. The dashed arrows indicate the occurrence of content switching (CW) during the closed-loop stimulation in Day 2 and 3. (**G**) SOL results of a 2-day nap protocol using Earable. Stimulation is used on Day 2 and not used on Day 1. (**H**) SOL results of 3-day full night sleep protocol using Earable. Stimulation is used on Day 2 and 3 and not used on Day 1.

We divided the investigation of the influence of audio stimulation on SOL into two phases. In the first phase, we tested the effectiveness of audio stimulation by using the gold standard PSG to capture the SOL. Additionally, we ran a nap session during the day instead of a full-night test. We played the audio through the two audio bone conductors integrated into our Earable headband to stimulate the subjects. We used the Earable headband in this experiment for conducting sounds only. The brainwave signals were collected using the gold-standard PSG hooked up to the subjects' heads. During this first phase, we turned off the audio after 50 minutes regardless of the subjects' sleep status. Fig. 7C shows the results of the two-day protocol where users were not subjected to the audio stimulation on the first day and then subjected to the stimulation on the second day. As seen in this figure, we successfully achieve a great reduction of SOL from the subjects when applying the audio stimulation. After testing on 166 subjects, the average SOL is reduced from $40.3 \pm 11.0$ to $16.2 \pm 11.1$ minutes, indicating the effectiveness of our stimulation protocol. To prove that the reduction of SOL in our case is not due to the "first-time" effect, we further conduct the PSG 3-day nap protocol and PSG 4-day nap protocol experiments. In the 3-day nap experiment, users are not subjected to audio stimulation on the first day but to audio stimulation on the second and third days. In contrast, in the 4-day nap experiment, users are not subjected to audio stimulation on the first and second days but subjected to audio stimulation on the third and fourth days. Figs. 7D and 7E demonstrate that applying audio stimulation reduces the SOL compared to the case without stimulation. For the 3-day nap protocol, the average SOLs over 28 subjects are $36.4 \pm 10.5$, $14 \pm 10.7$, and $18.8 \pm 17.2$ minutes for Day 1, Day 2, and Day 3, respectively. For the 4-day nap protocol, the average SOLs over 36 subjects are $41.2 \pm 11.0$, $38.9 \pm 10.5$, $22.1 \pm 13.9$, and $19.5 \pm 13.4$ minutes, respectively.

Having proven that our auditory stimulation effectively reduces SOL and thus helps users sleep faster, in the second phase, we proceeded to fully integrate the audio stimulation into our Earable headband to build the real-time closed-loop feedback sleep faster stimulation. Instead of always turning off the stimulation audio after 50 minutes of stimulation, our Earable headband turns off the audio right after the automatic detection of the SOL in our closed-loop stimulation. In addition, we also introduce the content switching (CW) functionality where the BM can be automatically changed if the subjects are not falling asleep as fast as within the first 20 minutes of their sleep session. Here, we introduce the Probability of Being Asleep (PoAs) parameter, which evaluates the level of "sleepiness" by approximating the probability of falling asleep given recently observed physiological data. Hence, if the PoAs slope is not high enough (i.e., the user is not falling asleep fast enough), our stimulation mechanism will automatically change the BM through the CW function. Fig. 7F depicts a PoAs example of a subject who spent three nights using our headband: the first night without stimulation and the second and third nights with closed-loop audio stimulation. Regardless of stimulation condition, the PoAs profiles usually consist of 3 parts: First, the PoAs rises relatively quickly within several minutes, probably because the user started to close eyes and relax. Then, the PoAs becomes relatively constant, indicating the user's period of trying to sleep. Finally, the PoAs increases and reaches another plateau value typically above 80%, coinciding with the SOL detected by our sleep staging algorithm as indicated by the solid arrows in Fig. 7F. Therefore, the PoAs metric we propose is a powerful manner of measuring the sleep wellness of users. As mentioned above, the CW functionality is introduced into our closed-loop audio stimulation when the PoAs slope is not high enough, as indicated by the dashed arrows. Upon introducing the closed-loop audio stimulation on Day 2 and Day 3, we observe that the PoAs profiles can reach the final plateau value faster than without stimulation. In this figure, the user's detailed SOLs in Day 1, Day 2, and Day 3 were 35, 25, and 25.5 minutes, respectively, suggesting the effectiveness of our stimulation.

To prove that the effectiveness of our stimulation is statistically significant, we further repeated the measurement on multiple subjects. Figs. 7G and 7H illustrate their results in cases of naps and full-night sessions, respectively. Specifically, Fig. 7G shows that our Earable headband successfully reduces the SOL from $36.6 \pm 10.0$ minutes (Day 1, without stimulation) to $16.9 \pm 12.0$ minutes (Day 2, closed-loop stimulation) after testing on 57 subjects in the nap testing protocol. In the full night home testing environment (Fig. 7H), we also show that the headband successfully reduces the SOL from $41.3 \pm 15.8$ minutes (Day 1, without stimulation) to $22.3 \pm 19.1$ minutes (Day 2, closed-loop stimulation) and $26.8 \pm 21.7$ minutes (Day 3, closed-loop stimulation) upon testing on 18 subjects.

**User experience survey**

During the development and evaluation process, we have also conducted a survey on users' experience with Earable to determine the gaps in our device and the real-life expectation. The questionnaire asks the users about their perspective of Earable's convenience, comfort level, and ease of use and how they are satisfied with the audio stimulation. This information provides us insight into improving the Earable system in the future.

Fig. 8 shows 12 questions we asked and the statistics on our participants' answers. In Fig. 8A, more than 95% of the participants reported that the band fitted well with their heads. For their first time using our device, it is very positive that more than 69% and 86% of the participants feel either neutral or comfortable while lying down on their side and back, respectively, with the Earable. Additionally, the band could stay in place in more than 75% full-night sleep studies in which the participants freely moved over the night. However, only about 69% of the participants felt the band did not hurt their heads after waking up. Besides that, nearly a half of them were woken up during the night because of the band. We believe that after repetitively using the device, people would be more comfortable and have a minimum effect caused by the pain as they can adjust the band tightness at a more appropriate level.

On the other hand, in Fig. 8B, we asked our participants about their experience with our own-built audio stimulation. Specifically, more than 87% of them like the audio. An equivalent number of participants felt comfortable and relaxed while listening to the audio. When we asked the participants to self-evaluate their duration of falling asleep, more than 45% of them imprecisely thought that they could sleep after less than 20 minutes. All of them started their sleep in longer than 20 minutes. Then more than 72% of the participants believe that our audio can help them go to sleep more quickly and easily. Finally, as our audio combines different components, we asked about their favourite components. As expected, the three key components

(i.e., guided breathing voice (GBV), a relaxation therapy voice (RTV), and background music (BM)) contribute significantly to the efficacy of the stimulation so that it can help to lull the participants into sleep better. Moreover, from this question, we observe that the appropriate audio volume is a decent factor contributing to the success of our fast sleep program. This information in Question 6 helps us know the need to improve the effectiveness of our current stimulation, build more options in the content switching feature, and develop a new feature on auto-volume adjustment based on their current sleep stage.

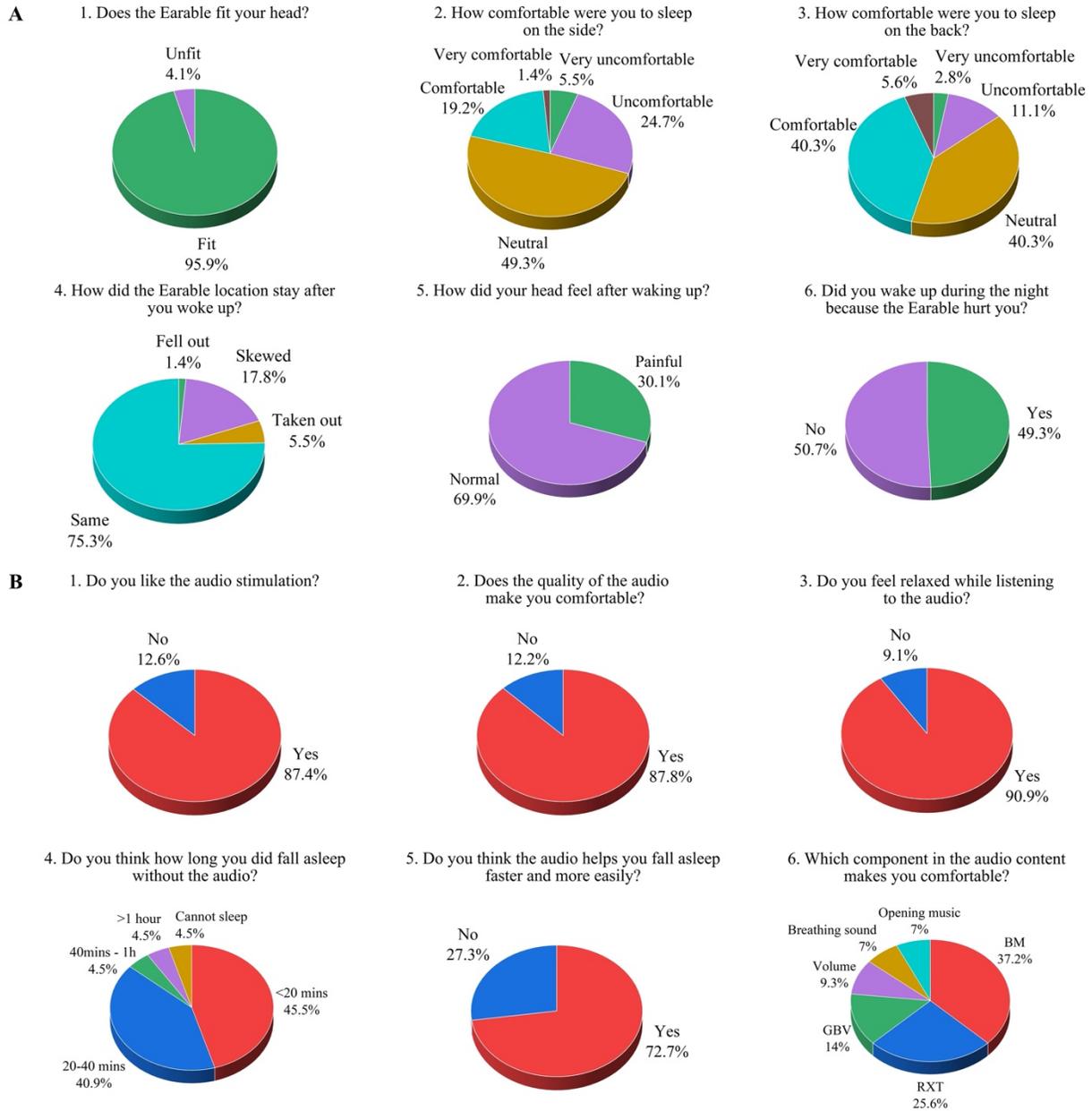

**Fig. 8. User experience questionnaire results.** (**A**) The perspective of convenience, comfort level, and ease of use. (**B**) The perspective of stimulation satisfaction.

# DISCUSSION

In this work, we developed a novel head-based sleep-aid system and demonstrated its use for high-fidelity recording, accurate monitoring, and effective stimulation of neural circuits for fast sleep falling in a large-scale study across multiple protocols. First, the data reported here shows that the electrophysiological signals traditionally used in sleep medicine can be captured using an optimally lightweight hardware and dry electrodes. With a substantial correlation with the PSG, the signals captured from the Earable system allow trained sleep technicians to identify typical sleep hallmarks such as Alpha rhythm, spindles, k-complexes, slow oscillation peaks, etc., which are critical for determining sleep stages and further providing insights into sleep quality. Clearly, the positions of physiological electrodes are reasonable and responsive to measure the brainwaves reflected from all over sources in the cortex. Moreover, wearing the Earable system does not need skin preparation, which is usually required in all epidermal recordings because skin impedance is highly variable between subjects when no skin treatment is used (*51*).

Second, the data shows that our set of algorithms to infer breathing rate, heart rate, and sleep posture using an accelerometer and a PPG sensor can achieve a great agreement with the gold standard. Conclusively, placing the 3-axis accelerometer in front of the forehead is a sufficiently sensitive position to detect miniature movements caused by other body parts from afar and large body movements. Similarly, the high agreement for heart rate proves that positioning the PPG sensor against the forehead can supply a reliable HR monitoring. However, we were unable to provide an oxygen saturation ($SpO_2$) measure in the recording due to subtly unstable placement of the band on the head. Further improvement of Earable in the design should solve this issue to enrich the amount of given data.

Third, we showed that Earable was powerful to perform real-time sleep staging using the data collected by the device with an accuracy in the comparable range of individual sleep technicians using PSG data. We introduced a methodology of using physiological signals to primarily guide the sleep staging process and smartly switching to PPG and acceleration data to temporarily estimate the sleep stage in occasions of noisy physiological sources. As a result, Earable can significantly reduce the number of unscored epochs with the reasonable accuracy in the range of PSG scoring consensus, which strengthens our precise delivery of stimulation in real-time. Finally, the data reported in Fig. 7 highlights the positive response to our own-built auditory stimulation for shortening the sleep onset latency of participants with sleeplessness. The fact that more than three-fourths of our subjects fell asleep in a shorter time is encouraging. Especially, our varied baseline design in a couple of the sub-studies gives more confidence that these effects are likely caused by the intervention (and not a first night effect or regression to the mean). Here, non-responders to our audio tended to be subjects with a high sleep time on the previous day reported prior to the study, suggesting a ceiling effect with key parameters integrated into the audio. Beyond this direct way of evaluating the efficacy of stimulation based on inter-scorer hypnograms, we further provide sleep variables as alternative macro-metrics statistically computed from the hypnogram to compare sleep characteristics with and without the stimulation in Table S1-S2. Gathering the results from both the figure and tables, the auditory stimulation that we have built and used appears to be effective and promising. However, various limitations of this study should be noted as well. First of all, it would be ideal to have a longer intervening washout period and increase the randomization of auditory intervention over more nights. Secondly, the sample was somewhat homogeneous in age, which about two-thirds of our subjects are in their 20s and 30s, and in national origin. A larger sample of more diverse sleepers in age and nation origin would have provided more generalizability to the general population. Lastly, some of our subjects reported hearing the crackling noise from the audio due to their high sensitivity to sound. Hence, further improvement of study protocols, subjects' demographics, and audio quality in the production should be conducted.

Taken together, all results indicate that the Earable headband holds great potentials for advanced sleep care monitoring and improvement at home. Future directions include evaluating and improving the capability, reliability, and stability of Earable for long-term recording and multi-purpose use (e.g., in movable scenarios), and optimizing the design and electrode materials to maximize the contact quality and the comfort level. This opens exciting possibilities for several exploratory studies in both clinical diagnostics and neuroscience research, such as those combining simultaneous electrophysiology and vital monitoring with various non-invasive brain stimulations to further boost the flexibility and effective aiding of the system for not only sleep but other mental health issues as well.

## MATERIALS AND METHODS

### Study protocol

We conducted two separate studies to prove the accuracy of our sleep staging model and the effectiveness of our auditory stimulation. Both were approved by the Institutional Review Board (IRB) of the Vietnam Society of Sleep Medicine in Protocol No. 01.10.2020-HĐĐĐ/HYHGNVN. We recruited participants without regard to gender or ethnicity from the local community through advertisement flyers. To confirm eligibility for the studies, volunteers first completed an online questionnaire of a detailed demographic, medical, health, sleep, and lifestyle. Next, we ruled out the volunteers with medical conditions, including sleep disorders (except the difficulty of falling asleep) and severe cardiac, neurological, or psychiatric comorbidity. Additionally, none of them had participated in any shiftwork or had traveled across more than one-time zone within the previous four weeks. We then randomly assigned the eligible participants to either of the studies. Finally, they gave written consent before attending their experimental session. After the sleep session, the technician removed all devices, and participants were debriefed and completed another questionnaire to identify any adverse events and provide their feedback. In addition, all participants received financial compensation commensurate with the burden of study involvement.

***Study I***: The purpose of this study is three folds: (1) to assess the reliability in recording electrophysiological and vital signals of Earable's sensors, (2) to develop the real-time sleep scoring model, and (3) to evaluate its accuracy. One hundred fifty-five participants between the ages of 19 and 33 were recruited. Each participant provided one to three nights of data based on their availability. Each night, a sleep technician sets up both PSG and the Earable to undergo an overnight sleep study at our sleep lab. The final analysis data set consisted of 155-full-night records from 84 female and 71 male participants. The PSG and Earable data recordings were synchronized a posteriori so that the records were well-aligned.

***Study II***: This study aims to objectively assess the efficacy of our proposed acoustic stimulation that could promote fast sleep by successfully reducing the SOL. There were multiple study phases designed to investigate the audio stimulation's influence on SOL and give the confidence that these effects are likely caused by our intervention and not a first night effect or other reasons. Three hundred and five participants between the ages of 19 and 55 were recruited. Each participant provided two, three, or four naps or three full nights of data. They were asked to refrain from caffeine and alcohol during experiment days. One day before the first study and during the whole experiment, subjects were asked to maintain a regular sleep and wake pattern, with a daily nocturnal sleep of 7 hours.

- *Phase 1*: we utilized daytime naps to run our studies with the data collected using the PSG. Specifically, the audio was played through speakers integrated into our headband. During the first 20 minutes, the participant watched a short video while the technician hooked up all the required electrodes to the participant's head. All lights were on, and windows were open to keep the participant awake. In the next 50 minutes, the lights were switched off, and the windows were closed so that the participant could begin the sleep session. After 50 minutes of sleeping, the participant was woken up, and the study ended.
- *Phase 2*: we kept running our nap studies using only the Earable headband. Hence, the participants went through a similar process described above at our sleep lab.
- *Phase 3*: we evaluated the stimulation efficacy through full-night sleep studies using only the Earable headband. To mimic the real-world experience for the participants, we gave them the headband to use at their homes, and their SOL with and without stimulation could still be monitored.

The final analysis data set consisted of 674-nap and 54-full night records from 213 female and 92 male participants.

**Ground truth collection and labeling of sleep data**

The PSG assessment was performed using an Alice 6 LDx Diagnostic sleep system (Philips, USA) with the following EEG derivations: F3/A2, F4/A1, C3/A2, C4/A1, O1/A2, O2/A1, LOC/A2, and ROC/A1; 200Hz sampling rate with a 0.3-35 Hz bandpass filter; bilateral electrooculographic (EOG), and bilateral chin electromyographic (EMG) recordings were performed. Airflow, thoracic movements, and fingertip oxygen saturation were also monitored. EEG gold cup-electrodes were attached to the participants' scalp with Ten20 conductive paste (Weaver and Company, Aurora, CO, USA), according to the international 10-20 system for electrode placement.

Inter-technician labeling contradictions are relatively common when comparing the scoring results from two or more technicians for the same sleep study. Although the AASM establishes guidelines for how to read and interpret sleep studies, disagreements naturally arise as a result of a given technician's interpretation of a section of sleep data. In a previous study in which 2,500 technicians were asked to score 1,800 epochs from 9 sleep sessions, the inter-technician agreement typically was 82.6% (*52*). Furthermore, even though sleep sessions are scored in discrete epochs, transitions between sleep stages are continuous events that may span more than one epoch. During this transition period, sleep hallmarks from more than one sleep stage may also be observable. Technicians must use their best judgment to score sleep epochs in the presence of these uncertainties, and differences in opinion between technicians can yield inconsistencies in the resulting hypnograms.

Building a set of ground truth labels based on the scoring of one technician can therefore bias the machine learning model to learn scoring patterns specific to this technician. To encourage the model to learn generalizable features from sleep data, a consensus hypnogram is constructed for each sleep study using the scoring hypnograms from at least 3 trained sleep technicians. Each technician performed sleep staging using PSG data that was simultaneously acquired during the sleep study to better assure the accuracy of our ground truth labels. In this approach, the ground truth sleep stage for each epoch is determined to be the maximally voted stage for that epoch by the set of technicians. When a tie exists (for example, if 3 technicians all scored the same epoch differently), then the epoch is scored according to the technician with the most frequent inter-technician agreement for this study (*53*).

**Earable sleep scoring algorithm**

Traditionally, according to the AASM Manual for the Scoring of Sleep and Associated Events (*40*), sleep studies are partitioned into epochs (non-overlapping, 30-second windows) and each epoch is assigned to one of 5 sleep stages, including Wake, N1 (NREM 1), N2 (NREM 2), N3 (NREM 3, or "Deep Sleep"), and R (REM). The assignment of an epoch to a sleep stage is based on principles of sleep architecture and the recognition of specific patterns and sleep hallmarks in the EEG, EOG, and EMG signals (*40*). Clinical sleep scoring is performed offline. That is, the data acquired in a sleep study is processed and scored by a sleep technician after completion of the sleep session, allowing the technician to reference both historical and future data when scoring a given epoch for a more accurate interpretation of the subject's sleep architecture. The Earable headband, on the other hand, performs real-time sleep analysis and sleep-enhancing closed-loop stimulation. Therefore, we develop the Earable sleep scoring algorithm in the manner of performing epoch-by-epoch real-time sleep scoring. Furthermore, in contrast to the 5-stage scoring methodology suggested by the AASM, the Earable sleep scoring algorithm performs 4-stage scoring, clustering stages NREM 1 and NREM 2 into a higher-level abstraction, Light Sleep. Motivating this sleep stage aggregation is the low inter-technician agreement when scoring sleep stages NREM 1 and NREM 2 and the similar biological phenomena that are observed in these lighter stages of sleep.

The Earable sleep staging algorithm is composed of 5 primary sub-modules - i) a Channel Selection module which identifies which data channels of the device contain signals with high enough fidelity for sleep staging, ii) a Dynamic Re-referencing module which adjusts the default channel referencing scheme depending on the channels that are inferred to be of good signal quality from the Channel Selection module, iii) a Data Preprocessing and Feature Extraction module which cleans and summarizes epoch data through the results of digital signal processing and statistical analyses, iv) a primary machine learning (PML) model which infers a probability distribution of sleep stages using the features computed from EEG, EOG, and EMG signals processed in the former module, and v) a secondary machine learning (SML) model designed to estimate a probability distribution of sleep stages using the features computed from PPG and IMU sensors, provided that electrophysiological signal data has been unstable from all electrode channels for a sufficient amount of time.

**SUPPLEMENTARY MATERIALS**

Materials and Methods

Tables S1 and S2

**Acknowledgments:** We sincerely thank Professor Simon Kyle and Rachel Sharman (University of Oxford) for consulting us for the study protocol.

**Funding:** This work was supported by the National Science Foundation grant no. CNS-1846541 (TV) and the Alfred P. Sloan Research Fellowship no. FG-2020-13110 (TV).

**Author contributions:** A.N. and T.V. conceived and designed the research and conceptualized the Earable device; A.N. and G.P. developed the scientific and algorithmic system; A.N., N.B., N.P., H.T., B.D., and T.V. produced the Earable device, developed the fabrication process, and performed electrical and electrochemical tests; A.N. and H.T. designed and performed preliminary effectiveness test protocols; A.N. and L.N. designed the study protocols and performed the experiments; S.D.Q provided clinical relevance context and feedback on auditory stimuli and study protocol design; A.N. and H.H.N. labelled the ground-truth data, A.N., G.P., B.D., and H.H.N. analyzed the data; S.H. and T.V. provided scientific and technical feedback on this work; A.N., G.P., B.D., N.B., H.T., and S.H. wrote the paper; All authors reviewed and commented on the manuscript.

**Competing interests:** Authors declare that they have no competing interests.

**Data and materials availability:** All data associated with this study are available in the main text or the supplementary materials. Additional information may be requested from A.N. (anh.nguyen@umontana.edu)

# Supplementary Materials for
# A Large-Scale Study of a Sleep Tracking and Improving Device with Closed-loop and Personalized Real-time Acoustic Stimulation

**Module 1: Channel selection**

EEG, EOG, and EMG signal qualities are highly dependent on the contact quality of a given electrode with the user's scalp. A body movement throughout the night may cause displacement of the headband on their head, disrupting the contact of one or more electrodes, and therefore causing the signal quality to become unsatisfactory for sleep scoring from these channels. Hence, we develop the Channel Selection algorithm as a lightweight Decision Tree classifier (*54*) which takes as input summary signal quality features from a single signal channel at the current epoch and outputs a binary result indicating the signal quality of the channel at this epoch (i.e., acceptable for scoring or unacceptable for scoring). This signal quality classification is performed for each of the six default headband channels to establish a set of "scorable" channels.

Alternatively, we apply rule-based channel selection sub-modules to establish the fidelity of the three PPG signals for SML model inference. Besides that, an outlier detection method is then applied to reject samples of IMU that are associated with irregular sensor movement and readings. These sub-modules together establish the utility of PPG and IMU data at the current epoch for sleep scoring.

**Module 2: Dynamic re-referencing**

Signal referencing is performed for EEG, EOG, and EMG signals in order to obtain signals with higher signal-to-noise ratios and fewer artifacts. By default, all channels are referenced to the CMS electrode of the headband. If either of the BE electrodes of the headband is found to have acceptable signal quality (as evaluated by the Channel Selection module), then these channels are introduced as new reference channels. Specifically, if the BE_L channel has acceptable signal quality, the right-hand side channels (FH_R, OTE_R, and BE_R) are re-referenced to BE_L (provided that they were deemed scorable by the Channel Selection module). This same re-referencing process is performed analogously for the left-hand side channels (FH_L, OTE_L, and BE_L) if and only if the BE_R channel has been inferred to have acceptable signal quality.

**Module 3: Data preprocessing and feature extraction**

Subsequent to Channel Selection and Dynamic re-referencing, a set of channels is available for sleep staging inference. Each of these channels is pre-processed through signal clipping to limit the amplitude range to [-500,500]uV and the application of a set of notch filters to remove powerline noise in the raw signal. Filters designed to isolate EEG, EOG, and EMG components from each signal are then applied to obtain surrogate EEG, EOG, and EMG signal data from each channel. Two PML model input tensors are finally computed from each channel. A spectrogram is computed from the EEG signal component and a 38-dimensional feature vector is computed to summarize time and frequency domain analyses from each channel's EEG, EOG, and EMG components.

**Module 4: Primary machine learning (PML) EEG/EOG/EMG-based model**

Morphological aspects of biosignal data critical to sleep stage inference and interpretation are present in the time, frequency, and time-frequency domains. Hence, we employ both recurrent and convolutional neural sub-networks in a hybrid input manner in this module to leverage such spatial and time-dependent features. Specifically, the PML model takes inputs as the spectrogram and the 38-dimensional feature vector for each epoch. The spectrogram is then fed into a shallow, 2-layer convolutional neural sub-network. On the other hand, the feature vector, in addition to 7 epochs of historical feature data, is fed to the recurrent neural sub-network to obtain an additional feature mapping. Output vectors from each subnetwork are concatenated to achieve a 928-dimensional latent feature vector which is finally presented to a single-layer dense classification head. Finally, we apply a final softmax layer (*55*) to the 4-unit output to achieve a probability distribution over the 4 sleep stage classes (sleep stages Wake, Light Sleep, Deep Sleep, and REM). At each epoch, this spectrogram and feature vectors are computed for each channel that has the acceptable signal quality and a forward pass through the network is performed using these data representations. If multiple channels have acceptable signal quality, the sleep stage probability distributions from each channel will be averaged to obtain an ensembled sleep stage confidence estimate. Finally, the estimated sleep stage for the epoch is the sleep stage associated with the highest confidence estimate.

The PML model architecture is tuned in the context of k-fold cross-validation (*56*) and designed to enable an optimal balance of size and efficiency in order to perform accurate, real-time sleep scoring on a user's mobile device. Model parameters were learned using a training/validation dataset composed of 106 randomly selected sleep studies (68% of the 155 total sleep studies). We then withheld the remaining 49 sleep studies (32%) for evaluation purposes. However, a primary challenge in developing algorithms for performing inference based on electrophysiological signals is the inter-user generalization. Many components of the EEG, EOG, and EMG signals vary substantially across people which makes the development of models that do not overfit the data it was trained on challenging. Furthermore, signal morphology is influenced by the hardware used for data acquisition, reducing the efficacy of simply training a model trained on large datasets acquired using clinical PSG hardware. Consequently, we perform a two-step training process of pre-training and then fine-tuning to mitigate the performance issues that can arise from these challenges.

In the pre-training phase of the PML model training, we train the network using sleep data acquired from clinical-grade PSG devices. It is because the signal hallmarks that sleep technicians use to stage sleep are most visible in this gold-standard data, as compared to data acquired from typical wearable devices, due to the fact that the hardware is carefully configured by the technician and is consistently monitored throughout the night and adjusted as needed. By pre-training the PML model on this PSG data from subjects of varying demographics (age, gender, etc.), our model learns signal features that are critical for scoring and generalization across users. While training, we stochastically optimize the model parameters using the Adam optimizer (*57*), iteratively minimizing categorical cross-entropy loss. Early stopping was used to terminate pre-training by monitoring classification accuracy on a withheld validation dataset (*58*).

Naturally, the distribution of spectrogram and feature vector values that are present in the Earable headband data will differ from those present in the clinical grade PSG data due to the different hardware configurations. As such, the pre-trained PML model must be tuned in accordance with this different data distribution. In this tuning process, we freeze the layers of the convolutional subnetwork while updating the layers of the recurrent sub-network and classification head by training on sleep data acquired using the Earable headband. Similar to the pre-training phase, we apply the Adam optimization in this iterative process. Moreover, we use a lowered learning rate and label-smoothing (*59*) of ground truth consensus labels to stabilize the optimization process in the presence of epochs with noisy data that do not contain data reflective of the ground truth sleep stage.

**Module 5: Secondary machine learning (SML) PPG and IMU-based model**

Although infrequent, it is possible for all electrodes to lose stable contact with the user's scalp during sleep. In this case, no scorable channel will be available after channel selection. As a result, no reliable EEG, EOG, and EMG information will be available for scoring. In this case, we employ the SML model which leverages features computed from the PPG and IMU sensors to infer the current sleep stage. Given that sleep stage transitions are typically apparent at more coarse time granularities from these data sources, as compared to in EEG, EOG, and EMG signals, this model estimates the current sleep stage of the user using two epochs (i.e., every minute) until reliable electrophysiological signals become available again.

Analogous to sleep staging with EEG, EOG, and EMG signals, the change in biological processes over time provides valuable information for estimating the current sleep stage. As such, we use a shallow recurrent neural network architecture in this SML model. The input to this model is a set of 24 features computed from PPG and IMU time-series data, including the estimated heart rate, respiratory rate, and movement information. In the development process of the Earable headband, the IMU and PPG sensors were only available for 27 of the total 155 sleep studies. Thus, we perform the model development, training, and validation using 18 (66.6%) of these studies, while the 9 (33.3%) remaining studies that were also present in the PML model test set were withheld for testing. Due to the limitation of data, we train this SML model in a typical, iterative method using the Adam optimizer.

**Function 1: Heart rate estimation from PPG signals**

Using a 5-second, non-overlapping window, we segment the signals collected from the three PPG channels (IR, Red, and Green) to estimate a user's heart rate. The signal segments of each channel that have high enough fidelity according to the PPG channel selection algorithm are bandpass filtered to isolate the AC components of the signal. However, the channels that do not pass this channel selection step are ignored for the current timestep. Then, the resulting set of filtered signals are normalized to have a zero mean and unit variance and aggregated by taking the mean of each filtered and normalized signal segment. Systolic peaks are detected in this resulting signal and heart rate estimates are computed using the durations between successive systolic peaks. We finally apply moving average smoothing to reduce the occurrence of sudden changes in heart rate estimation due to signal noise.

**Function 2: Respiratory rate estimation from IMU signals**

To estimate the respiratory rate of a user every minute, we use a 60-second, non-overlapping window of IMU data. Given the orientation of the IMU in the Earable headband, small movements resulting from inhalation and exhalation are detectable via the Y-axis data of the IMU. After applying a bandpass filter to this time-series data to isolate breathing frequency components, we detect respiratory oscillations via the peaks and troughs of the filtered data. The median duration between successive peaks in this 60-second window is finally computed to estimate the user's respiratory rate at this timestep.

**Function 3: Automatic content switching**

Users may respond differently to various sources of audio content when trying to fall asleep. While some may find the sounds of rainfall, for example, soothing and sleep-inducing, others may find it distracting. Hence, we develop an automatic content switching (ACS) algorithm that changes the background music in the audio stimulation in response to the real-time sleepiness trend of a given user. This algorithm helps users fall asleep smoothly, without interruption or delay caused by audio stimulation that feels disruptive for falling asleep. As the user falls asleep, their level of sleepiness is continuously inferred every epoch via the PoAs value. During the sleep onset process, if we observe a negative sleepiness trend occurring with the current source of background music, this music content will be automatically changed in order to better guide the user to sleep.

**Function 4: Audio content recommendation**

When it is time to select a new source of audio to play for the user (i.e., when the user starts a sleep session, after the current audio finished playing, or when prompted by the ACS algorithm) historical data from the user can be leveraged to select audio content that will optimally encourage sleep induction. The audio content recommendation (ACR) algorithm is an instantiation of the reinforcement learning Multi-Armed Bandit problem (*60–64*). A content recommendation agent attempts to choose audio content for the stimulation that will maximize the subject's PoAs rate of change (suggesting a maximal rate of increase in the subject's sleepiness level). Audio content preferences for each user are updated by observing rewards from the previous sleep session, where the reward for a given audio content is computed as the user's PoAs rate of change while that content was being played. The value of a given audio content is represented by a normal distribution, whose parameters are updated after each reward observation for that content. These distributions are specific to each user, as users will naturally have different audio preferences. Initially, without any prior knowledge, all audio content has a value distribution with large variance, as the preference of the user is unknown. Over time, as rewards are observed for a given audio content, the variance of the content's value distribution decreases as the preference model distribution becomes closer to the true expected reward for that content. The Thompson sampling algorithm (*65, 66*) is employed to update the preference model distribution parameters over time. When the content recommendation is needed, the value distribution of each audio content (excluding the content that was most recently used) is sampled and the content yielding the highest value sample is selected. This algorithm allows for custom user preferences without extensive prior user data. Additionally, new sources of audio content may be released at any time without altering any other content distribution parameters, as a new value distribution may simply be instantiated for this content.

**Table S1. Sleep variables as macro metrics computed on the hypnogram of PSG data collected in the three nap protocols**

|  | 2-Day Nap Protocol | | 3-Day Nap Protocol | | |
|---|---|---|---|---|---|
|  | Nap 1 | Nap 2 | Nap 1 | Nap 2 | Nap 3 |
|  | No Stimulation | Stimulation | No Stimulation | Stimulation | Stimulation |
| TIB (min) | 51.22 ± 3.75 | 51.20 ± 4.48 | 53.46 ± 8.19 | 52.77 ± 4.93 | 58.23 ± 7.80 |
| SOL (min) | 33.27 ± 8.52 | 14.92 ± 8.65 | 31.13 ± 6.66 | 12.57 ± 7.75 | 14.04 ± 8.97 |
| TST (min) | 7.80 ± 9.13 | 28.36 ± 12.48 | 14.82 ± 12.09 | 32.05 ± 13.19 | 31.89 ± 20.17 |
| SE (%) | 15.12 ± 17.39 | 55.27 ± 23.56 | 25.62 ± 20.34 | 61.10 ± 25.03 | 53.96 ± 31.93 |
| LS (min) | 6.21 ± 7.07 | 19.10 ± 9.59 | 7.96 ± 6.71 | 15.82 ± 7.77 | 13.46 ± 8.46 |
| LS (% of TIB) | 50.43 ± 45.95 | 69.85 ± 29.79 | 40.77 ± 36.86 | 53.00 ± 24.79 | 47.37 ± 30.20 |
| DS (min) | 1.51 ± 4.36 | 9.11 ± 10.64 | 6.86 ± 8.43 | 15.82 ± 10.63 | 17.13 ± 13.73 |
| DS (% of TIB) | 6.99 ± 18.90 | 25.52 ± 26.48 | 27.08 ± 30.31 | 42.47 ± 23.85 | 39.59 ± 28.48 |
| REM (min) | 0.08 ± 0.83 | 0.15 ± 0.75 | 0.00 ± 0.00 | 0.41 ± 1.25 | 1.30 ± 3.70 |
| REM (% of TIB) | 0.41 ± 3.91 | 0.41 ± 1.96 | 0.00 ± 0.00 | 0.97 ± 2.96 | 2.33 ± 6.15 |

|  | 4-Day Nap Protocol | | | |
|---|---|---|---|---|
|  | Nap 1 | Nap 2 | Nap 3 | Nap 4 |
|  | No Stimulation | No Stimulation | Stimulation | Stimulation |
| TIB (min) | 52.64 ± 6.09 | 51.81 ± 10.42 | 53.17 ± 5.28 | 53.96 ± 5.85 |
| SOL (min) | 34.63 ± 10.05 | 34.04 ± 7.20 | 18.86 ± 10.22 | 17.69 ± 10.85 |
| TST (min) | 7.32 ± 9.67 | 9.42 ± 9.56 | 24.90 ± 14.34 | 28.47 ± 13.15 |
| SE (%) | 12.96 ± 16.67 | 17.41 ± 17.50 | 46.45 ± 26.76 | 53.45 ± 25.16 |
| LS (min) | 5.51 ± 7.08 | 7.33 ± 6.97 | 18.32 ± 11.27 | 20.40 ± 10.43 |
| LS (% of TIB) | 45.77 ± 46.75 | 57.85 ± 44.28 | 69.17 ± 31.26 | 71.83 ± 28.11 |
| DS (min) | 1.81 ± 5.01 | 2.08 ± 4.49 | 6.40 ± 8.20 | 7.93 ± 8.31 |
| DS (% of TIB) | 7.01 ± 18.84 | 8.81 ± 18.06 | 19.04 ± 21.01 | 22.34 ± 22.62 |
| REM (min) | 0.00 ± 0.00 | 0.00 ± 0.00 | 0.18 ± 0.80 | 0.14 ± 0.82 |
| REM (% of TIB) | 0.00 ± 0.00 | 0.00 ± 0.00 | 0.68 ± 3.25 | 0.28 ± 1.63 |

**Table S2. Sleep variables as macro metrics computed on the hypnogram of Earable data collected in the nap and full-night study protocols**

|  | 2-Day Nap Protocol | | 3-Full Night Protocol | | |
| --- | --- | --- | --- | --- | --- |
|  | Nap 1 | Nap 2 | Night 1 | Night 2 | Night 3 |
|  | No Stimulation | Stimulation | No Stimulation | Stimulation | Stimulation |
| TIB (min) | 52.10 ± 6.79 | 50.96 ± 8.25 | 321.81 ± 125.86 | 372.22 ± 49.93 | 331.33 ± 107.43 |
| SOL (min) | 31.20 ± 8.77 | 16.99 ± 10.20 | 43.53 ± 20.64 | 29.00 ± 21.80 | 30.44 ± 23.21 |
| TST (min) | 9.53 ± 10.95 | 22.07 ± 14.26 | 244.89 ± 108.74 | 300.42 ± 84.29 | 248.83 ± 122.80 |
| SE (%) | 29.34 ± 33.73 | 71.93 ± 45.62 | 69.70 ± 23.06 | 80.05 ± 21.45 | 73.59 ± 28.40 |
| LS (min) | 7.23 ± 8.39 | 14.88 ± 10.46 | 128.67 ± 60.43 | 148.75 ± 43.73 | 127.53 ± 66.19 |
| LS (% of TIB) | 51.06 ± 44.15 | 63.88 ± 35.63 | 52.12 ± 18.51 | 47.47 ± 14.50 | 45.37 ± 20.33 |
| DS (min) | 2.30 ± 4.66 | 7.19 ± 9.08 | 80.92 ± 41.98 | 101.53 ± 40.87 | 78.92 ± 43.09 |
| DS (% of TIB) | 10.34 ± 19.43 | 22.08 ± 26.13 | 30.30 ± 14.87 | 31.53 ± 10.68 | 29.57 ± 15.99 |
| REM (min) | 0.00 ± 0.00 | 0.00 ± 0.00 | 35.31 ± 28.79 | 50.14 ± 26.89 | 42.39 ± 36.02 |
| REM (% of TIB) | 0.00 ± 0.00 | 0.00 ± 0.00 | 12.05 ± 9.17 | 15.44 ± 7.10 | 13.95 ± 9.34 |